\documentclass[journal = nalefd]{achemso}
\usepackage{textcomp}
\usepackage{color}
\usepackage{xr}
\usepackage{graphicx}

\author{Sharmila N. Shirodkar}
\email{sharmila.shirodkar@howard.edu}
\affiliation{Department of Physics and Astronomy, Howard University, Washington D.C. 20059, United States}
\author{Pratibha Dev}
\email{pratibha.dev@howard.edu}
\affiliation{Department of Physics and Astronomy, Howard University, Washington D.C. 20059, United States}
\title{Non-linear Hybrid Surface-defect States In Defective Bi$_2$Se$_3$}

\begin{document}

\begin{abstract}
Surface-states of topological insulators are assumed to be robust against non-magnetic defects in the crystal. 
However, recent theoretical models and experiments indicate that even non-magnetic defects can perturb these states. 
Our first-principles calculations demonstrate that the presence of Se vacancies in Bi$_2$Se$_3$,
has a greater impact than a mere n-doping of the structure, 
which would just shift the Fermi level relative to the Dirac point.
We observe the emergence of a non-linear band pinned near the Fermi level, 
while the Dirac cone shifts deeper into the valence band. 
We attribute these features in the bandstructure to the interaction between the surface and defect states,
with the resulting hybridization between these states itself depending on the position and symmetry of the Se vacancy relative to the surfaces.
Our results bring us a step closer to understanding the exotic physics emerging from defects in Bi$_2$Se$_3$ that remained unexplored in prior studies.
\end{abstract}

\section{Introduction}
Topological insulators are a new class of materials with a gapped bulk and metallic surfaces (for 3D materials) or edges (for 2D materials) \cite{kane2005,bernevig2006}.
These surface states are topologically non-trivial, and originate from the strong spin-orbit coupling that preserves time-reversal symmetry.
They are known to be robust against defects, strain and disorder, assuming that the perturbation is modest enough to preserve the original character of the material. 
Since these surface states exhibit Dirac fermion behavior (linear dispersion), not only are they being explored for applications in quantum computing and spintronics,
but their topological protection allows one to test and explore exotic and fundamental concepts in physics, such as Majorana particles \cite{grover2014}
and magnetic monopoles \cite{qi2009}.
As different proposed applications of topological insulators depend on the assumption of a linear Dirac cone,
it is critical to understand if the surface states indeed remain unperturbed, or if they are reshaped by the non-magnetic defects, and to what extent.
In fact, a recent theoretical model and experiments \cite{miao2018,zhong2017,teague2012} indicate that the latter is indeed the case.
However, earlier first-principles calculations \cite{felser,wang2020} of these defects did not consider the reshaping of the surface states due to their interactions with defect states.

Amongst the bismuth selenide family of compounds (Bi$_2$Se$_3$, Sb$_2$Te$_3$ and Bi$_2$Te$_3$),
Bi$_2$Se$_3$ \cite{xia2009,zhang2009} is the most widely studied 3D topological insulator, with a large bulk-band gap of 0.3 eV. 
Selenium vacancies (VSe) are known to be a common defect in Bi$_2$Se$_3$,
and are assumed to be responsible for the n-type doping of the as-grown crystals \cite{navratil2004,shaham2021}.
Due to high volatility of selenium \cite{kim2019}, formation of additional vacancies on the surface of cleaved Bi$_2$Se$_3$ slabs is expected.
This is possibly responsible for the observed aging of the slabs, as they show changes in electronic properties with time \cite{king2011,bianchi2010}.
Though a vast literature exists on the formation energy of Se vacancies in Bi$_2$Se$_3$ \cite{adam,west2012,felser,wang2020,Dai},
very little is discussed about the nature of the defect states, 
their placement in the electronic band structure and the effect of their positions on the properties of Bi$_2$Se$_3$ surface states.
Furthermore, most theoretical works have either considered very large densities of vacancies,
such as a surface termination with all top selenium atoms removed,
or symmetric defects on both surfaces that preserve the inversion symmetry of the slab \cite{felser}.
Hence, the effect(s) of lower concentrations of vacancies on Bi$_2$Se$_3$ remain unknown.
To address these important questions, we carry out a detailed analysis on the effect of Se vacancies on the Dirac cone of the Bi$_2$Se$_3$ slabs,
explore the coupling between the surface and defect bands, and their emergent signature in the band structure and STM images.

\section{Computational Details}
SIESTA (Spanish Initiative for Electronic Simulations with Thousands of Atoms) package \cite{siesta1,siesta2} was used to carry 
out the Density Functional Theoretical (DFT) calculations. A plane wave mesh cutoff of 300 Ry was used for non-spin polarized calculations,
whereas a lower cutoff of 150 Ry was used for calculations including spin-orbit interactions.
The electron-ion interactions were modeled with fully-relativistic norm-conserving pseudopotentials obtained from PseudoDojo \cite{pseudodojo,psml},
with exchange correlation energy between electrons approximated by the GGA functional parameterized by Perdew-Burke-Ernzerhof (PBE)\cite{perdew1996generalized}.
A double zeta polarized basis set was used in our calculations.
The Brillouin Zone (BZ) integrations were carried out over a Monkhorst Pack k-point mesh of size 6$\times$6$\times$6 for primitive bulk cell
and 6$\times$6$\times$1 for slab of size 1$\times$1$\times$nQL, where nQL stands for the number of quintuple layers (QL).
Defects were created in supercell slabs of sizes 6$\times$6$\times$6 QL and 5$\times$5$\times$7 QL.
Only BZ center integrations were carried out for the defective supercells.

The bulk primitive cell was relaxed (including spin-orbit coupling) till the Hellman-Feynman forces on the atoms were \textless 0.01 eV/\AA~
and the stress components in each direction were \textless 0.001 eV/ \AA$^3$.
This relaxed bulk structure was used to create 1$\times$1$\times$nQL (nQL = 6, 7) slabs, which was further relaxed (atoms only, with spin-orbit coupling)
until the Hellman-Feynman forces on the atoms were \textless 0.01 eV/\AA.
This slab structure was then periodically repeated in the in-plane lattice directions to generate the 6$\times$6$\times$6 QL (1080 atoms) 
and 5$\times$5$\times$7 QL (875 atoms) supercell slabs.
To estimate the structural relaxation due to Se vacancy defects, we carried out non-spin polarized calculations on the defective supercell slabs
and the atoms were allowed to relax till the Hellman-Feynman forces on the atoms were \textless 0.04 eV/\AA.
In addition, we kept the Bi atoms fixed in their respective planes (allowed to move in-plane), to fix the interplanar spacing of the layers to that of bulk.
Because we exclude the spin-orbit effects in the relaxation calculations, we keep the Bi atoms fixed in their respective planes (allowed to move in-plane)
since GGA parameterization overestimates the interplanar spacing (a known drawback). 
This constraint is reasonable since it mimicks experimental slabs, where the defect concentration is approximately 
two orders of magnitude lower than in our calculations depending on the quality of the sample, and would not affect the bulk interplanar spacing significantly.
A vacuum of 24 \AA~was included in the non-periodic direction to minimize the interaction between the periodic images.

\begin{figure}[!ht]
\centering
\includegraphics[width=12cm]{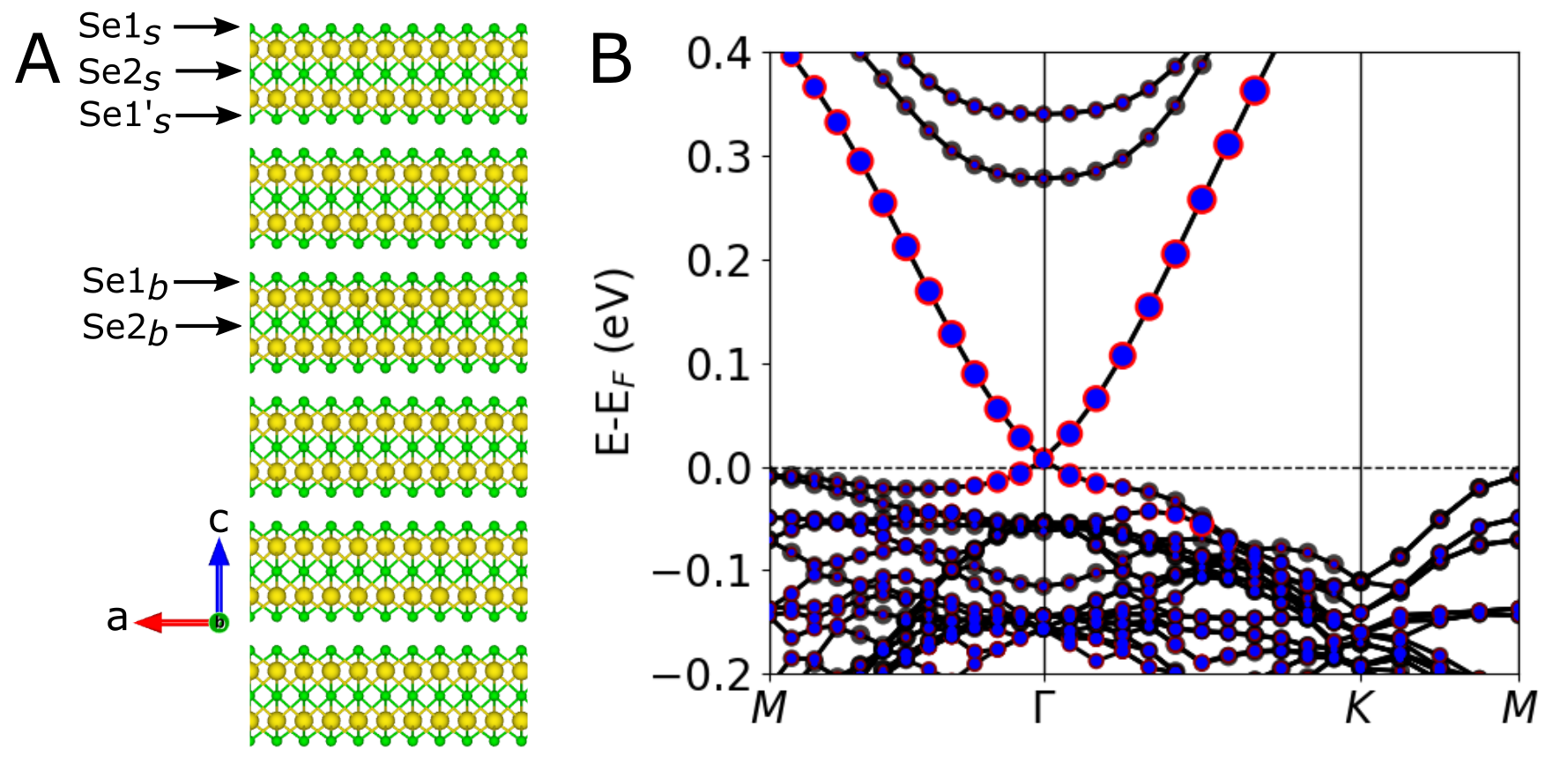}
\caption{Structural model (6$\times$6$\times$6 QLs slab) and electronic structure of pristine Bi$_2$Se$_3$.
A) Symmetry inequivalent positions (Se1, Se1\textquotesingle~and Se2) at which defects are created in the surface layer (subscript s) and in the bulk (subscript b). 
Note: the 6$\times$6$\times$6 QLs slab is separated from its periodic image in the c-direction by 24 \AA~of vacuum.
Se (Bi) atoms are shown by green (yellow) circles.
B) Electronic structure of pristine 6$\times$6$\times$6 QLs slab, showing the Dirac cone emerging from the topological surface states at the Fermi level.
The bands are colored to show contributions from different QLs to the electronic states (red= top QL, blue= bottom QL and black= bulk contributions).
The charge density plots of the wavefunctions are given in Figure S1.}
\label{666_pristine}
\end{figure}

\section{Results \& Discussion}
\subsection{Vacancy configurations}
Bulk Bi$_2$Se$_3$ has R-3m (\#166) space group symmetry with a) Bi at Wyckoff position 6c (0,0,0.3985), and
b) Se with two symmetry inequivalent Wyckoff positions, 3a (0,0,0) and 6c (0,0,0.2115); and lattice constants a = b = 4.19 \AA~and c = 28.63 \AA \cite{nakajima}.
It is a layered material with each layer containing 5 atoms (a formula unit), and therefore known as a quintuple layer (QL). 
The QLs have -A-B-C- stacking along the \textquoteleft z (c)\textquoteright~direction. 
The DFT estimates of lattice constants for the 1$\times$1$\times$nQL (nQL = 6 and 7) cell are a = b = 4.21 \AA~
i.e. overestimation by $\approx$ 0.4\% from the experimental value.
Whereas the estimated bulk lattice parameter, c = 28.92 \AA~is overestimated from the experimental value by 1\%, well within the limits of DFT errors. 
Due to the difference between the electronic topology of the bulk with respect to vacuum,
cleaving the bulk into a slab gives rise to topologically protected Dirac surface
states at the Fermi level, assuming the slab is thick enough to reduce interaction between top and bottom surfaces.
For slabs with nQL $\ge$ 6, we find that the surface states become almost degenerate (i.e. produce a Dirac cone) at the Fermi level 
with a negligible gap between the conduction and valence band Dirac cones ($<$ 1 meV.
Pertsova et al. \cite{pertsova2014} have shown that this gap essentially goes to zero for $>$ 40 QLs, which is when the top and bottom surface states become decoupled.

We construct 6$\times$6$\times$6 QLs and 5$\times$5$\times$7 QLs slabs to study the effects of Se vacancy defects on the surface states,
with the 2 slabs representing different in-plane defect densities.
In the 6$\times$6$\times$6 QLs slabs, we create single VSe defects that break the inversion symmetry of the slab, 
capturing the defect- and substrate-induced asymmetry in experimental samples\cite{park2013}.
On the other hand, in the 5$\times$5$\times$7 QLs slabs, we create pairs of VSe defects that conserve the inversion symmetry of the top and the bottom QL 
layers, allowing for a direct comparison with the existing literature\cite{felser}.
A comparison of results for the 6$\times$6$\times$6 QLs and 5$\times$5$\times$7 QLs slabs also allows us to 
highlight and differentiate between the effects of inversion symmetry breaking on the bandstructure and those due to defects themselves.
As the overall results for the two structures are qualitatively similar, those for the 5$\times$5$\times$7 QLs slab are discussed in the Supporting Information, S2.
In both the 6$\times$6$\times$6 QLs and 5$\times$5$\times$7 QLs slabs, we create Se vacancies
at the symmetry-inequivalent Wyckoff positions (3a and 6c) at the surface and in the interior (i.e. bulk) of the slab. 
Since the surface breaks the symmetry between the two Se atoms at 6c,
we distinguish them as Se1 on the surface and Se1\textquotesingle~below the surface (interface of topmost and next QL).
Se2 is the Se atom at the 3a site which is sandwiched between two BiSe layers (see Figure \ref{666_pristine}A).
We will henceforth refer to vacancies in the surface (bulk) QL with a subscript \textquoteleft s\textquoteright (\textquoteleft b\textquoteright).
An Se vacancy in a 6$\times$6$\times$6 QLs slab corresponds to a defect concentration of 
$\approx$ 10$^{19}$ cm$^{-3}$ (in-plane concentration = 2.6 $\times$ 10$^{13}$ cm$^{-2}$),
comparable to the values reported in experiments\cite{walsh,adam}, which range from 10$^{11}$ to 10$^{20}$ cm$^{-3}$.
We note that, previous theory works \cite{felser,wang2020} simulated a single Se vacancy per 3$\times$3 in-plane supercell,
while we use 6$\times$6 and 5$\times$5 in-plane supercells.
The small supercell size in the earlier works results in stronger interactions between the periodic images of the vacancies,
changing the effect of the defect states on the topological surface states of Bi$_2$Se$_3$.
The band structure of a pristine 6$\times$6$\times$6 QLs slab is shown in Figure \ref{666_pristine} B, and shows the Dirac cone at the Fermi level.

\subsection{Bulk vacancies}
\begin{figure*}[!ht]
\centering
\includegraphics[width=15cm]{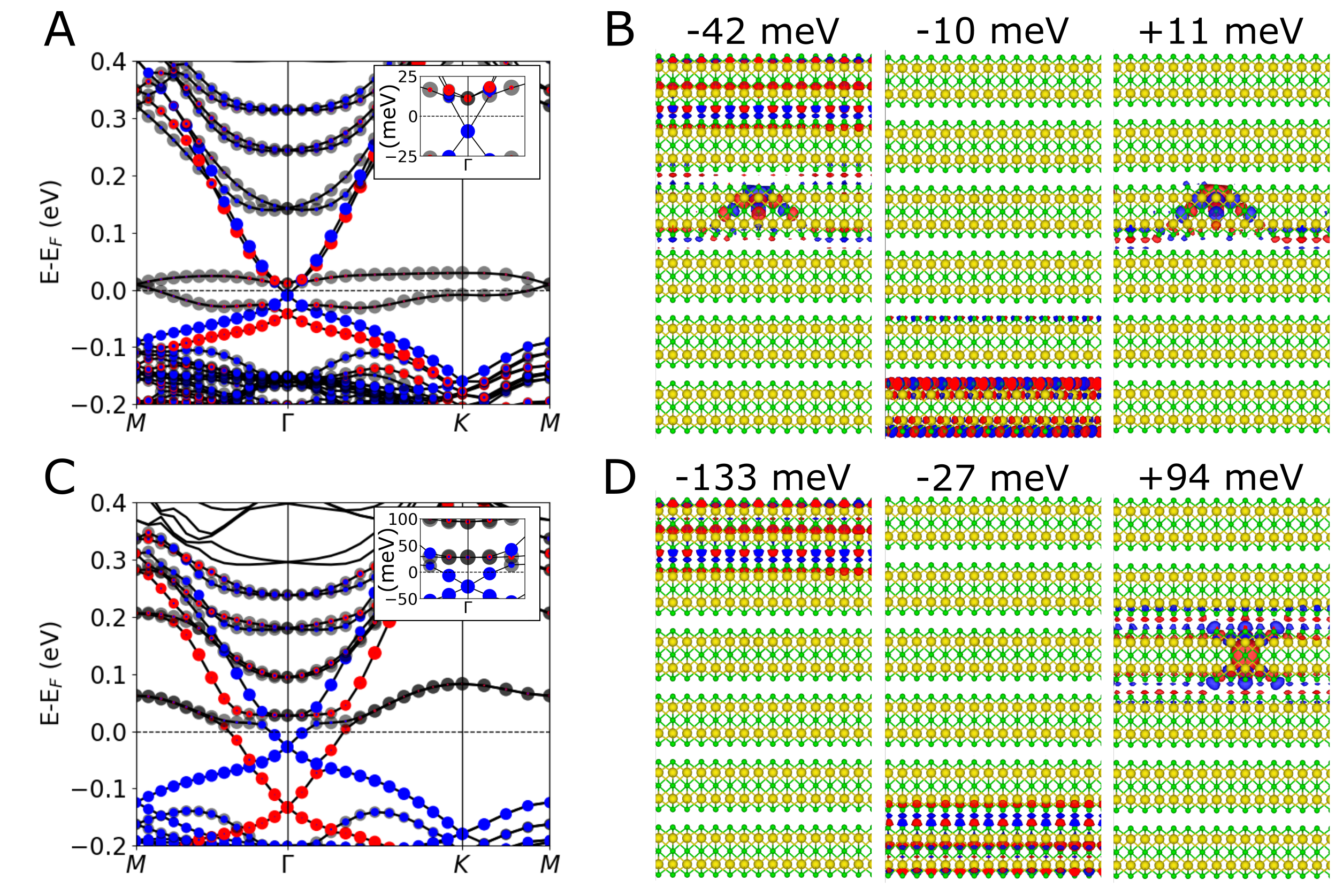}
\caption{Electronic band structure and wavefunctions (at the $\Gamma$-point) of bulk Se vacancies in a 6$\times$6$\times$6 QLs slab. 
Band structures of A) VSe1$_b$ and C) VSe2$_b$ are colored to show contributions from different QLs to the electronic states
(red= top QL, blue= bottom QL and black= bulk contributions).
The insets are the zoomed-in band structures at the Fermi level and the $\Gamma$-point in the BZ.
Note that the energy units in the insets are meV, as opposed to eV in the main plot.
The charge densities of the wavefunctions are shown in B) for VSe1$_b$ and D) VSe2$_b$.
Se (Bi) atoms are shown by green (yellow) circles. The opposite phases of the wavefunctions which are calculated at the $\Gamma$-point are denoted by red and blue colors in the charge density plots.}
\label{666_bulk}
\end{figure*}

We first explore how the vacancies in the bulk of the Bi$_2$Se$_3$ slab modify its electronic structure properties.
The bandstructures for the two inequivalent defects, VSe1$_b$ and VSe2$_b$, are plotted in Figures \ref{666_bulk} A and C, respectively.
The contributions of the top (bottom) surface (QL) to the electronic states are drawn with red (blue) circles; black circles represent the contribution of the bulk.
Note that the states at the $\Gamma$ and M points in the Brillouin Zone (BZ) are doubly degenerate, due to their topological nature. 
The defect breaks the structure inversion symmetry (SIA) of the slab \cite{shan2010}, with the top QL being closer to the defect.
This lifts the degeneracy of the surface states localized on the top and bottom surfaces, 
with the top surface state moving lower in energy due to its larger interaction with the defect.\cite{shan2010,park2013}
The Dirac cones associated with the top surface states are shifted by $\approx$ -42 meV and -133 meV below Fermi level for VSe1$_b$ and VSe2$_b$, respectively.
In contrast, for the inversion-symmetry conserving VSe2$_b$ in 5$\times$5$\times$7 QLs slab (see Figure S2 C),
both the Dirac cones are degenerate due to conservation of inversion symmetry.
The shift of the Dirac point below the Fermi level, which is due to interaction of the surface states with the defect \cite{miao2018,zhong2017,xu2017},
(discussed later in detail) likely depends on the distance of the defect from the surface,
and is expected to be reduced for defects that are much farther away from the surfaces in thicker slabs.

Figures \ref{666_bulk} B and D show the wavefunctions at the Dirac points (i.e. $\Gamma$-point) for the two bulk defects.
In the case of VSe1$_b$, we find that the topological character (linear dispersion) for the top and bottom surface states is preserved,
with the associated Dirac cones shifted below the Fermi level by about -42 meV and -10 meV (see Figure \ref{666_bulk} B, left-most and middle panels).
Figure \ref{666_bulk} B (right-most panel) also shows that the unoccupied conduction band minimum at $\Gamma$.
It is about +11 meV above the Fermi level, and is mostly localized around the defect site.
Interestingly, the wavefunctions at about -42 meV and a lower isosurface plot of the state at about +11 meV
show that they are actually admixtures of the top surface states and defect states, being bonding and anti-bonding states, respectively.
In the case of the bulk defect, VSe2$_b$, the top surface state is shifted by $\approx$ -133 meV,
whereas the bottom surface state at $\approx$ -27 meV is localized to the bottommost QL.
Unlike the hybridization between surface and defect states seen for VSe1$_b$,
the defect-state associated with VSe2$_b$ is decoupled from the surface states (see Figure \ref{666_bulk} D at -133 meV and -27 meV).
Anti-crossing of the surface states and the defect band (see Figure \ref{666_bulk} C at $\approx$ +0.05 eV) confirms their decoupling.
This shows that although the VSe2$_b$ defect is at a similar distance from both the surfaces as VSe1$_b$,
the Wyckoff position of the Se vacancy in bulk, affects the electronic structure properties of a Bi$_2$Se$_3$-slab differently.
In part, these differences might stem from the fact that a VSe2 defect preserves the intra-layer inversion symmetry as opposed to VSe1, which breaks it.
We find similar behavior in the 5$\times$5$\times$7 QLs slab with VSe2$_b$ (see Figure S3). 

\begin{figure*}[!ht]
\centering
\includegraphics[width=15cm]{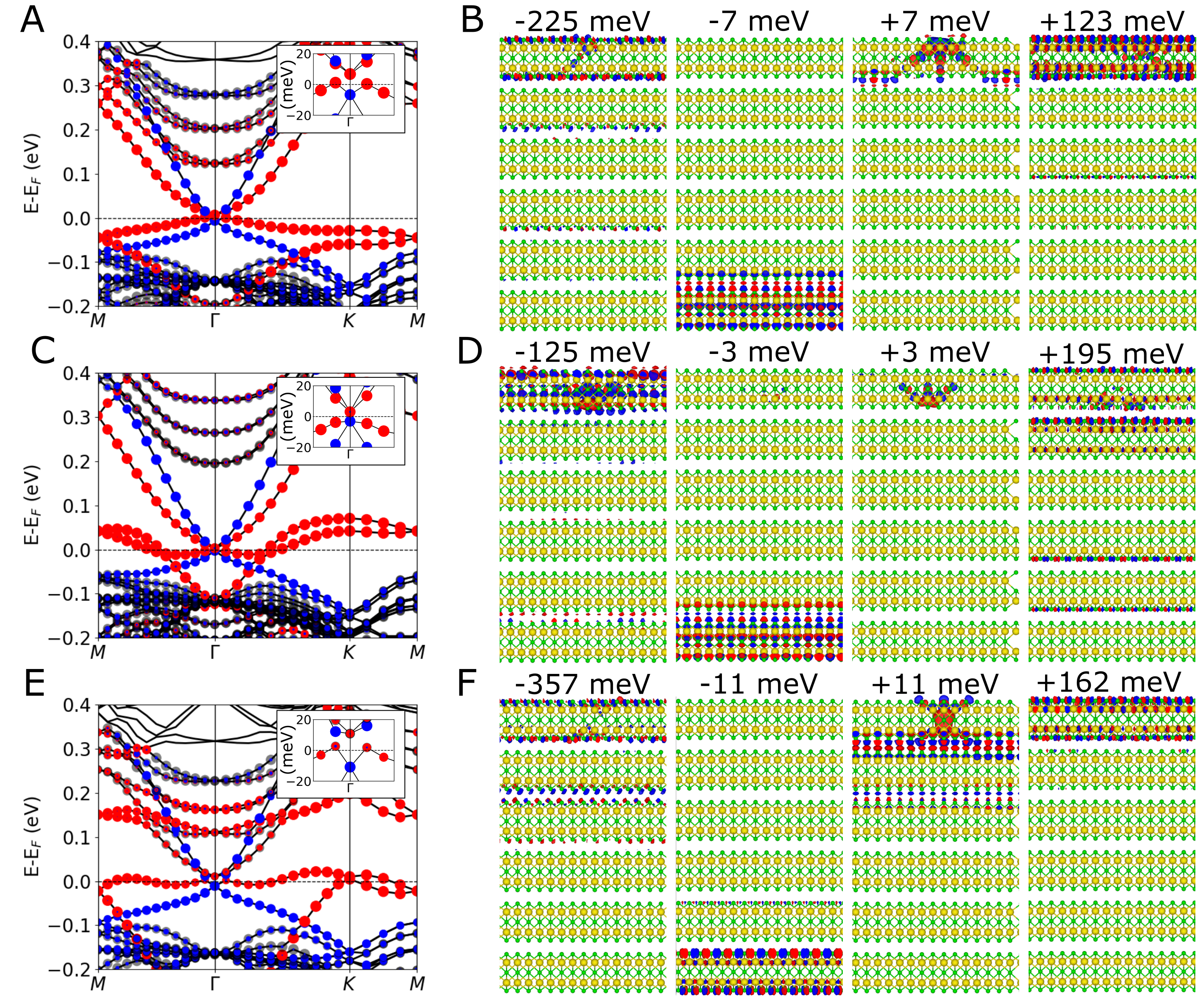}
\caption{Electronic band structure and wavefunctions (at the $\Gamma$-point) of surface Se vacancies in a 6$\times$6$\times$6 QLs slab. 
Band structures of A) VSe1$_s$, C) VSe1$_s$\textquotesingle~and E) VSe2$_s$, are colored to show contributions from different QLs to the electronic states
(red= top QL, blue= bottom QL and black= bulk contributions)
The insets are the zoomed-in band structures at the Fermi level and the $\Gamma$-point in the BZ.
Note that the energy units in the insets are meV, as opposed to eV in the main plot.
The charge densities of the wavefunctions are shown in B) for VSe1$_s$, D) for VSe1$_s$\textquotesingle~and F) for VSe2$_s$.
Se (Bi) atoms are shown by green (yellow) circles. The opposite phases of the wavefunctions which are calculated at the $\Gamma$-point are denoted by red and blue colors in the charge density plots.}
\label{666_surface}
\end{figure*}

\subsection{Surface vacancies}
Unlike bulk Se vacancies, where the interaction between the defect and surface states can be reduced by considering thicker slabs,
surface vacancies merely by virtue of their position are inherently coupled to the surface states.
For the surface defects VSe1$_s$, VSe1$_s$\textquotesingle~and VSe2$_s$,
the calculated band structure clearly shows strong interactions with the topological surface states (see Figures \ref{666_surface} A, C and E),
and the emergence of a non-linear (almost parabolic) band at the Fermi level (that was missing in for bulk vacancies).
In addition, the top QL state shifts much deeper into the valence band as compared
to that in the presence of bulk defects (see Figure S6 for band structure plotted for wider energy range).
Following the signatures of the top QL contribution to the band structure (colored red in Figure \ref{666_surface}, left column) into the valence band,
we identify these surface states by plotting the charge densities of their wavefunctions
at $\Gamma$-point (see left-most column in Figure \ref{666_surface} B, D and F).
The surface states of the defective top QL are shifted by $\approx$ -225 meV, -125 meV and -357 meV 
for VSe1$_s$, VSe1$_s$\textquotesingle~and VSe2$_s$, respectively. 
Not only are they localized to the topmost QL, but they also exhibit linear dispersion in the band structure (see Figure S6),
with the same Dirac cone features as in a pristine slab (see Figure \ref{666_pristine} B).
In each of these Dirac cones that are found much deeper in the valence band, the lower branch of the cone shows the expected upward dispersion, 
which arises from the $k^5$ term in the Dresselhaus spin-orbit coupling \cite{basak2011}.
This further confirms the topological protection of surface states in Bi$_2$Se$_3$,
even in the presence of a considerably large in-plane concentration of Se vacancies at the surface.

Furthermore, we see a shift of the Dirac cone from the bottom (pristine) layer (see Figure \ref{666_surface}, 
band structure in blue and second panel of the wavefunctions) below the Fermi level
by -7, -3 and -11 meV for VSe1$_s$, VSe1$_s$\textquotesingle, VSe2$_s$, respectively.
The downward shift from the Fermi level is a result of the inter-surface coupling between the defect and bottom QL \cite{shan2010},.
This is confirmed from the smaller shift of -5 meV for a single surface vacancy in the 5$\times$5$\times$7 QLs slab (see Figure S5,
and an equal and opposite shift seen in the hybrid, mostly defect-type, state around $\Gamma$-point (Figures \ref{666_surface} B, D and F, third panel).
We expect the shift to reduce with increasing slab thickness.
Interestingly, we also find that the top part of the surface state cone from the top QL surface has shifted into the conduction band by  
$\approx$ 123 and 162 meV for VSe1$_s$ and VSe2$_s$, respectively (Figures \ref{666_surface} B, D and F, right-most panel).
On the other hand, the surface state at $\approx$ 195 meV for VSe1$_s$\textquotesingle~shows greater delocalization,
with significant contributions from the next QL due to hybridization with the bulk conduction band states.
The splitting of the top and bottom part of the surface state Dirac cone agrees with the work of Black-Schraffer et al.\cite{annika2012}.

\subsection{Origin of the defect band}
\begin{figure}[!ht]
\centering
\includegraphics[width=8cm]{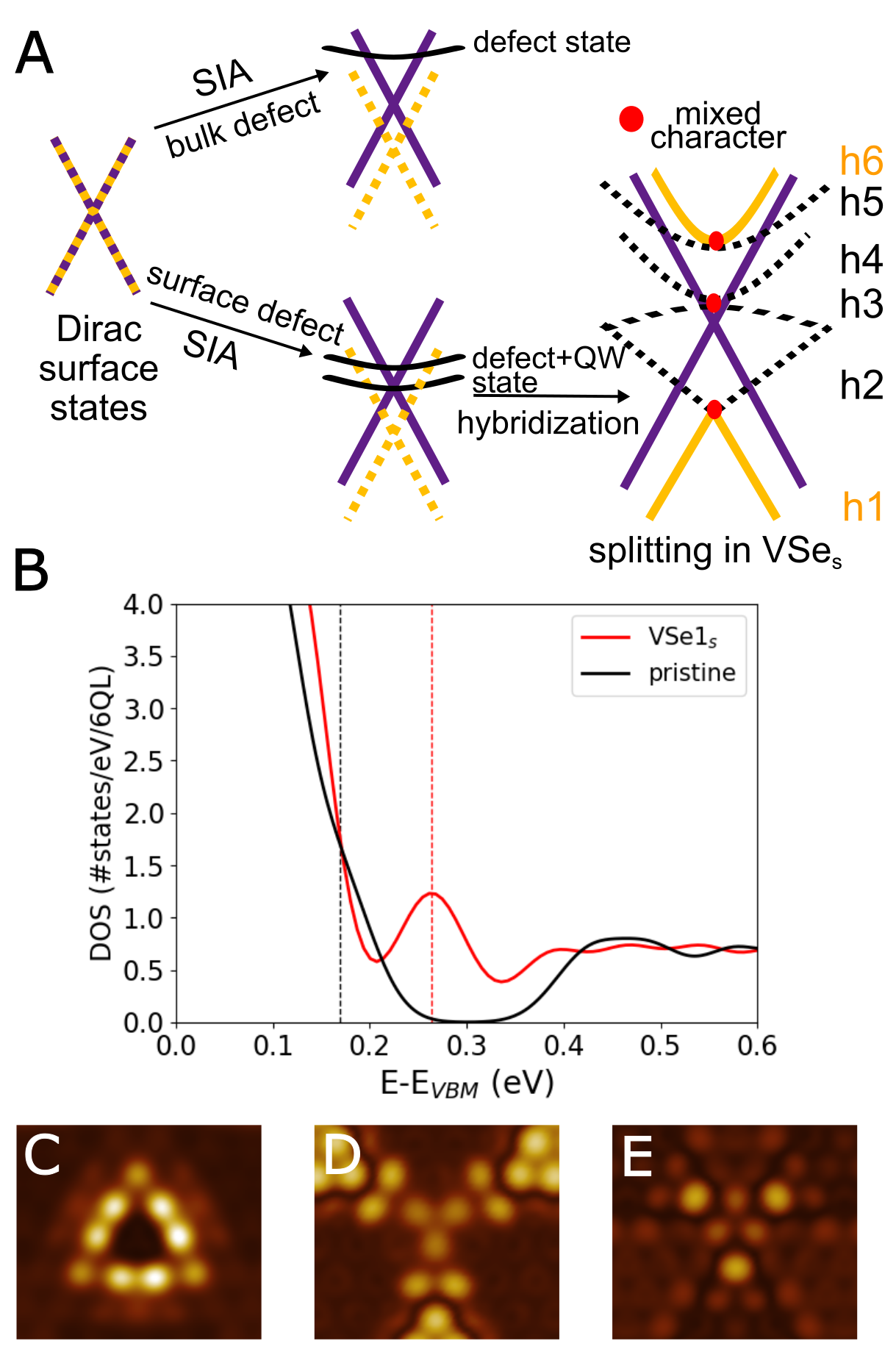}
\caption{Defect band and its origin. A) Energy diagram postulating the origin of hybrid bands (h1$-$h4),
due to interaction between Se vacancies and surface states.
SIA stands for structural inversion asymmetry.
The defect states are shown in black, and the top and bottom QL Dirac cones are shown by yellow dashed and purple solid lines, respectively.
B) Comparison of the density of states (DOS) of the pristine 6 QL Bi$_2$Se$_3$ slab (black) and VSe1$_s$ (red).
The Fermi levels are marked by black (pristine) and red (defect) dashed lines.
Constant height STM images of C) VSe1$_s$, D) VSe1$_s$\textquotesingle~and E) VSe2$_s$, for V= +0.3 V (sampling the conduction band).
VSe1$_s$ can be distinguished by a dark spot in its STM image, VSe1$_s$\textquotesingle~by a pattern resembling the nuclear radiation symbol,
and VSe2$_s$ by 3 bright spots.}
\label{666_stm}
\end{figure}

In addition to identifying the Dirac points associated with the top and bottom surface states,
we investigate the origin of bands lying between $\approx$ -0.2 eV to 0.2 eV (see Figures \ref{666_surface} A, C and E),
focusing on the non-linear bands near the Fermi level in this range.
Figure \ref{666_stm} A is a schematic diagram depicting our proposed origin of these bands.
The Se vacancy within the top QL introduces SIA,\cite{shan2010}
which lifts the degeneracy between the top (dashed yellow in Figure \ref{666_stm} A) and bottom QL layer (solid purple) Dirac cones.
The Se1 and Se1\textquotesingle~positions possess 3m point group symmetry, whereas the Se2 position has -3m symmetry.
Here we focus our discussion on the VSe1$_s$ defect, but similar explanation is valid for VSe2$_s$ and VSe1$_s$\textquotesingle~as well.
Due to the 3m symmetry of the VSe1s defect, the dangling \textquotesingle p\textquotesingle~orbitals of the Bi atoms (near the vacancy)
split as the singly degenerate A1 and doubly degenerate E bands.
The A1 defect band is near the Fermi level,
whereas the E bands are higher in the conduction band at $\approx$ 0.8 eV (see Figure S9).
On closer inspection of the conduction bands (see Figure \ref{666_surface} A, C and E), we find states in the bulk band gap
that were absent for the pristine slab (see Figure \ref{666_pristine} B).
These states are the Quantum Well (QW) states that are induced by the band bending at vacuum and defective QL interface
due to the symmetry breaking and doping of the top QL by the defect \cite{park2013}.

In the absence of any interaction between the defect and the surface states, the band structure would resemble Figure \ref{666_stm} A (lower-branch, middle plot).
Interaction between the defect/QW and topologically protected top surface states \cite{seixas2015,annika2012} creates hybrid bands (h1$-$h6),
shown in Figure \ref{666_stm} A, lower-branch, right-most plot (also see S7 for a numerical model based on \cite{seixas2015}).
Here, the defect band spin-splits into two branches, one of which mixes with the lower part of the Dirac cone of the top QL.
The Dirac cone that is deep in the valence band is a hybrid (h1$+$h2) with the lower part (h1) originating from the top QL surface state,
and the upper part (h2) from the defect.
This mixing is evident in the charge densities of the wavefunctions (see Figures \ref{666_surface} B, D, F at $\Gamma$ point).
h3 is the other spin-split defect band that primarily interacts with the lower part of the Dirac cone.
Since the defect forms an integral part of the surface, h1, h2 and h3 also acquire some defect and surface characters, respectively.
Similarly, the upper part of the original Dirac cone (of the top QL) mixes with the QW state \cite{park2013} present above the
Fermi level in the bulk gap region and forms, h4, h5 and h6 bands.
Here, the non-linear/parabolic band (h4) above the Fermi level, is mostly defect-like in character (see Figure S10).
This mixing behavior observed here is quite different from an earlier work by Seixas et al., \cite{seixas2015}
where a \textquoteleft twinning\textquoteright~of the Dirac cone at the interface of a topological insulator and a trivial insulator was reported.
In that case, the mixed states only emerge at the high symmetry points in the BZ.
Hence, the hybridization, which results in the non-linear/parabolic band, h4,
described in our work is a unique signature characterizing the Se vacancies in Bi$_2$Se$_3$.
h1$-$h6 also exhibit helical spin-texture (see Figure S11),
being admixtures of the topologically protected surface states and defect states. Lastly, we note the interaction between the defect and the bottom QL.
This shifts the defect state and the bottom surface state above and below the Fermi level, respectively.
The magnitude of the shift appears to be correlated to: a) how deep the h1$+$h2 band has shifted into the valence band,
and b) screening of the vacancy by adjacent layers. The shift is larger for VSe2$_s$ in comparison with VSe1$_s$ and VSe1$_s$\textquotesingle,
as the h1$+$h2 hybrid band is also lower (see Figures \ref{666_surface} A, C and E).
In the case of VSe1$_s$\textquotesingle, though it has similar placement in the QL as VSe1$_s$,
the screening effect of the adjacent QL for the VSe1$_s$\textquotesingle~makes the splitting negligibly small (see Figure S12). 
A quantitative analysis of the extent of the defect and surface state splitting (h1, h6 and h3, h4 separation) has been explored
in a generalized theoretical model of a strong three-dimensional topological insulator by Black-Schraffer et al. \cite{annika2012},
where they show that the magnitude of the shift of the surface states into the conduction and valence bands is dependent on the strength of the defect potential.

The h1$-$h6 bands preserve the topological character of the surface states,
i.e. they span the bulk energy gap and connect the bulk valence and conduction bands.
They are similar to the so-called \textquoteleft topological dangling bond states,\textquoteright\cite{lin2013,felser}
which were reported when an entire layer of Se atoms is missing from the top QL.
It is important to note that the presence of surface states (h1 and h6) outside the bulk band gap region
encourages scattering from the bulk states. Thus, the surface states lose their protection from spin backscattering processes.
The dispersion of h2 and h3 for VSe1$_s$\textquotesingle~and VSe2$_s$
(between energies of -0.05 to 0.05 eV, Figures \ref{666_surface} C and E),
gives rise to multiple Fermi surface nestings, which is similar to Bi$_2$Se$_3$ slabs cleaved along different planes \cite{lin2013}.
This feature allows for spin backscattering processes, previously forbidden in a pristine slab. 
For $|$energies$|$ $\geq$ 0.05 eV, the multiple Fermi-surface nesting disappears and the spin backscattering processes again become forbidden.
Thus, by controlling the position of the Fermi level via gating, one can control the surface spin current in defective Bi$_2$Se$_3$.
This possibility opens up new applications in spintronics that can replace the existing semiconductor based devices.

It is important to note that even though the parabolic band, h4 (see Figures \ref{666_surface} A, C and E), 
looks like a 2DEG (2D electron gas) \cite{king2011,bianchi2010} in Bi$_2$Se$_3$, its origin and spin texture reveal the opposite.
h4 shows spin-momentum locking with a slight out-of-plane \textquoteleft z\textquoteright~component
(due to defects \cite{jin2012}, see Figure  S11), in contrast with a 2DEG which is spin-degenerate \cite{king2011,bianchi2010}.
A 2DEG band in the topological insulators is known to originate from band bending induced by defects\cite{hsieh2008} 
or by change in the van der Waals gap between layers \cite{chulkov2011}, not hybridization.
It is also formed well above the Dirac point \cite{hsieh2008,king2011,bianchi2010,chulkov2011} in the conduction band.
These differences lead us to believe that h4 is a previously unreported characteristic of Se vacancies in Bi$_2$Se$_3$. 
The signature of the defect is also seen as a resonant defect state above the Fermi level 
in the density of states (see Figure \ref{666_stm} B), consistent with predictions and observations \cite{teague2012,miao2018,zhong2017,xu2017} in earlier works.
 
Our calculations for defect in the 6$\times$6$\times$6 QLs also confirm experimental reports of
a Rashba-like splitting of the conduction band states,\cite{king2011} (see Figures \ref{666_bulk}, \ref{666_surface} and
S13). This splitting is due to inversion asymmetry introduced by the asymmetric defects in 6$\times$6$\times$6 QLs
slabs, that is absent in the symmetrically placed defects in 5$\times$5$\times$7 QLs slabs (see Figure S2).
The exception is the VSe1$_s$\textquotesingle~defect, where this splitting is negligible.
Although an exact explanation of the latter is outside the scope of this work,
different contributing factors might include a countering of the Rashba-like splitting by the
effects of local strain (due to the defect) on the bandstructure, or the screening of the defect potential (see Figure S12).
Figures \ref{666_stm} C-E are simulated STM images (at constant height and V = +0.3 eV) for surface defects, 
providing their experimental signatures.
We find that our STM images are in reasonable agreement with experimental observations of Dai et al.\cite{Dai}.
The differences could be due to sampling of a different energy range in our calculations.

\section{Conclusions}
Our detailed study of Se vacancies in Bi$_2$Se$_3$ shows that a neutral Se vacancy state hybridizes with the
bottom part of the surface state Dirac cone and shifts it deeper into the valence band, i.e. it n-dopes the layer. 
On the other hand, the top part of the Dirac cone also hybridizes with the quantum well (bulk) state.
The QW-defect-surface states interaction, in turn, gives rise to a non-linear (parabolic) band with primarily defect character
at the Fermi level, while shifting the top Dirac cone into the conduction band.
We find that the non-linear bands at the Fermi level originating at the surface prohibit spin-backscattering processes for VSe1$_s$,
just like the surface states in pristine Bi$_2$Se$_3$ slabs.
Hence, in this structure, the directionality of the spin current at the surface is preserved.
The origin, spin texture and position with respect to the Fermi level of this band is unlike the 2DEG observed in Bi$_2$Se$_3$
\cite{king2011,bianchi2010}, which makes it a unique feature of surface vacancies in this material.
For the first time, our DFT results confirm the presence of such a defect band consistent with earlier reports \cite{miao2018,zhong2017}.
The unique dispersion of this band allows one to switch the spin backscattering processes \textquoteleft on/off\textquoteright~by controlling its occupancy,
providing additional tunability to the spin current. Deterministic placement of Se-vacancies can provide an added functionality
to topological-insulator based applications in spintronics. This would be a considerable improvement over traditional semiconductor spintronic materials.
Studying these defects under stress, strain, electric fields and gas contaminants would be an exciting future scope of this work.

\section{Supporting Information}
The Supporting Information includes results for symmetric Se vacancies in 5$\times$5$\times$7 QLs slab,
numerical model to explain the hybridization of defect states with the topological surface states, Rashba splitting in the electronic structure,
spin-momentum locking plots, and the convergence studies of the electronic structure with plane wave cutoff, displacement of Bi atoms and Hamiltonian matrix elements.

\section{Acknowledgements}
We would like to thank Dr. Ivan Naumov for helpful discussions and inputs, and Dr. J. R. Knab for help with copy editing.
This work was supported W. M. Keck Research Foundation grant and NSF Grant number DMR-1752840.
We acknowledge the computational support provided by the Extreme Science and Engineering Discovery Environment (XSEDE) under Project PHY180014,
which is supported by National Science Foundation grant number ACI-1548562, and Maryland Advanced Research Computing Center.

\bibliography{References} 

\clearpage

\begin{figure}[!]
\centering
\includegraphics[width=8.25cm]{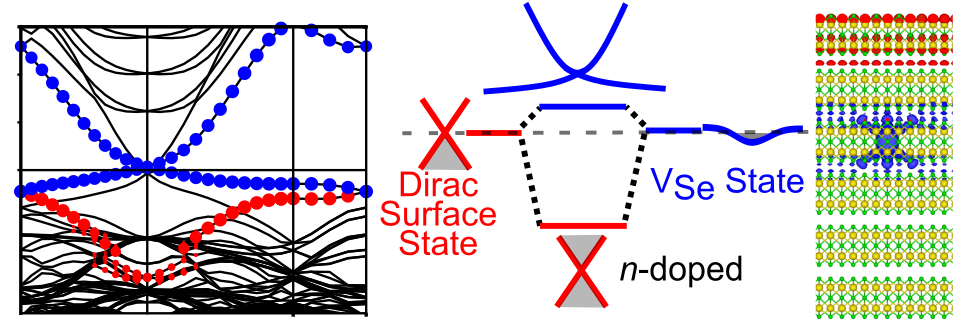}
\caption{Table of Contents Graphic.}
\end{figure}
\end{document}